\documentclass[newfonts=false,format=sigconf,10pt,letterpaper]{acmart}
\usepackage{times}
\usepackage{geometry}
\usepackage[normalem]{ulem}

\usepackage{times}  
\usepackage{hyperref}
\usepackage{enumitem}
\usepackage{amsmath,xcolor}
\usepackage{graphicx}
\usepackage{subcaption}
\usepackage{color}

\hypersetup{pdfstartview=FitH,pdfpagelayout=SinglePage}

\usepackage{algorithm}
\usepackage[normalem]{ulem}
\usepackage{url}
\usepackage{enumitem}
\usepackage{tikz}
\usepackage{caption}
\usepackage{subcaption}
\usepackage{xspace}

\usepackage{amsmath,xcolor}

\usepackage{array,ragged2e}

\usepackage{booktabs}
\usepackage{multirow}
\usepackage{csquotes}
\usepackage{amsmath}
\usepackage{blindtext}
\usepackage{graphicx}
\usepackage{epsfig,endnotes}
\usepackage[T1]{fontenc}
\usepackage[noend]{algpseudocode}
\setcopyright{none}
\renewcommand\footnotetextcopyrightpermission[1]{} 

\newcommand{\ballnumber}[1]{\tikz[baseline=(myanchor.base)] \node[circle,fill=.,inner sep=1pt] (myanchor) {\color{-.}\bfseries\footnotesize #1};}

\usepackage{tikz}

\makeatletter
\def\BState{\State\hskip-\ALG@thistlm}
\makeatother

\newenvironment{packeditemize}{
	\begin{list}{$\bullet$}{
			\setlength{\itemsep}{1.5pt}
			\setlength{\labelwidth}{8pt}
    		\setlength{\leftmargin}{10pt}
			\setlength{\labelsep}{3pt}
			\setlength{\listparindent}{\parindent}
			\setlength{\parsep}{1.5pt}
			\setlength{\parskip}{1.5pt}
			\setlength{\topsep}{1.5pt}}}{\end{list}}

\newcommand{\mavericknospace}[0]{\maverick}
\newcommand{\maverick}[0]{{M{\small AVERICK}}\xspace}

\newcommand{\blue}[1]{\textcolor{black}{#1}}

\usepackage{verbatim}

\hypersetup{draft}

\newif\iffullpaper
\fullpapertrue

\begin{document}\sloppy

\title{MAVERICK: Proactively detecting network control plane bugs using structural outlierness}

\author{Vasudevan Nagendra}
\affiliation{\institution{Stony Brook University}}
\email{vnagendra@cs.stonybrook.edu}

\author{Abhishek Pokala}
\affiliation{\institution{Stony Brook University}}
\email{apokala@cs.stonybrook.edu}

\author{Arani Bhattacharya}
\affiliation{\institution{KTH Royal Institute of Technology}}
\email{aranib@kth.se}

\author{Samir Das}
\affiliation{\institution{Stony Brook University}}
\email{samir@cs.stonybrook.edu}

\begin{abstract}
Proactive detection of network configuration bugs is important to ensure its proper functioning and reduce cost of network administrator.
In this research, we propose to build the control plane verification engine \mavericknospace that detects the bugs in the network control plane i.e., network device configurations and control plane states. \maverick automatically infers {\em signatures} for the control plane configurations (e.g., ACLs, route-maps, route-policies and so on) and states that allows administrators to automatically detect bugs with minimal human intervention. \maverick achieves this by effectively leveraging any {\em structural deviation} i.e., {\em outliers} in the network configurations that is organized as simple or complexly nested key-value pairs. 
The outliers that are calculated using {\em signature-based outlier detection} mechanism are further characterized for its severity and ranked or re-prioritized according to their criticality. We consider a wide set of heuristics and domain expertise factors for effectively to reduce both false positives and false negatives. 

Our evaluation on four medium to large-scale enterprise networks show that \maverick can automatically detect the bugs present in the network with $\approx$75\% accuracy. Furthermore, With minimal administrator input i.e., with a few minutes of signature re-tuning, \maverick allows the administrators to effectively detect $\approx$94 -- 100\%  of the bugs present in the network, thereby ranking down less severe bugs and removing false positives.



\end{abstract}

\maketitle
\pagestyle{plain}



\section{Introduction}\label{sec:introduction_hotnets}
In general, network downtime of a medium-scale enterprise network costs around \$140K -- \$500K per hour, for which the human errors acts as the key contributing factor~\cite{gartner_downtime_cost,ugly_truth_downtime_cost}. The fundamental goal of network management and downtime mitigation is {\em proactive detection} of the control plane bugs and ability to quickly troubleshoot the errors that occurred due to human errors and mis-configurations. Today network administrators either relies on custom {\em home-made} scripts 
`or' model checking-based verification tools for analyzing the network configurations to detect specific types of bugs in the network (e.g., reachability analysis, routing issues, failure impact analysis and so on) ~\cite{arc,fogel2015_generalapproach_batfish,panda2015new_directions,khurshid2013veriflow}. Such tools provide limited bug detection capability i.e., does not provide comprehensive coverage about the list of bugs present in the network configurations. 

Therefore, a generic control plane bug detection 
engine that proactively detects comprehensive list of bugs present in the network and allows administrators to supply their domain expertise to fine-tune bug detection is essential in effectively securing or protecting the networks from down times and vulnerabilities. Performing the aforementioned bug detection task with minimal administrator's intervention is key aspect that highlights the efficacy of the tool. 

Traditionally, the bug detection can be efficiently achieved by defining unique signature to each of the network property and matching each of the configuration instance with the 
signature. For example, an ACL that allows web traffic from LAN network to Internet needs to be specified on to a group of network devices along the path of the traffic until the traffic reaches the border gateway of enterprise network. Therefore, multiple devices should have either same or similar ACL. Deviation from the actual ACL definition will be considered as bug. As a similar example, route maps are used for defining the set of route entries that are required to be redistributed to target routing process, requiring the route maps to be specified on to multiple routers. 

Effectively identifying such signatures (or specifications) and manually providing comprehensive list of signatures for detecting bugs is a daunting task. But, not providing {\em signatures (i.e., about what specifically needs to be looked for in the network configurations)} results in bugs and errors that go undetected and results in false negatives that plague the soundness of any bug detection tool. In our observation there are legions of bugs that remain undetected even with networks ``vetted'' by checking tools, because of a lack of capability that allows the signature to be specified and used for bug detection. 

Therefore, the current network verification tools falls short along following key dimensions: ($i$) proactively detecting control plane bugs (e.g.,  human errors and configuration mistakes) without (or with minimal) administrator's intervention, ($ii$) ability to effectively incorporate domain expertise in fine-tuning the bug detection, ($iii$) automatically inferring policies or signatures from the network configurations that allows administrators and tools to effectively verify the sanity of configurations, while providing comprehensive bug detection coverage, ($iv$) generalize findings i.e., signatures or policies inferred from one network and apply it to other networks or organizations, and ($v$) finally, surfacing the bugs that are critical allowing administrators to channelize their energy in addressing critical bugs rather than wasting time on false negatives.


To address the above challenges, we propose \mavericknospace, an agile network verification tool that exploits structural deviations (i.e., outlierness\footnote{Outlierness is the deviation of the network configurations from its general population or most popular values.}) among the network configurations for detecting the bugs. The key enabler of \maverick is its ability to automatically infer signatures from the network configurations, which is used for efficiently detecting the bugs present in the network, without false negatives. 
\maverick also incorporates inputs from the administrators allowing the tool to fine-tune the detection precision. In addition, \maverick also proposes the need for generalization by which the signatures that are developed for an network can be used with for other networks.

We improve the accuracy of our bug detection mechanism and efficiently re-prioritize the bugs to surface them to administrators on the basis of their severity. We calculate severity of the bugs using following key metrics such as feature importance (i.e., network structural properties such as ACLs, route-maps, IPSec tunnel configurations and so on), feature dependency, the locality of the configuration on specific node, outlierness score from the similarity with signatures, and customized page ranking used for ranking bugs. These metrics allows \maverick to effectively prioritize bugs on the basis of their severity pushing the false positives or less critical bugs to the bottom of the list.
We prove the efficacy of \maverick by showing that it provides a mean precision of 78\% without administrator input, and 94\% using a few minutes of administrator input on a set of four medium to large-scale networks.


\noindent In summary, our paper makes following key contributions:

\begin{packeditemize}

\item
We illustrate the limitations of existing techniques with examples and motivate the need for signature-based bug detection mechanisms based on outliers (see \S \ref{sec:problems_existing_approaches_hotnets}). 

\item
We highlight the techniques we explored to automatically infer signatures for various properties of network present in the network configurations using their structural outlierness to detect control plane bugs. 
We highlight key metrics we explored to enhance the usability of our framework by calculating severity and ranking of bugs to reprioritize them and reduce false positives (see \S \ref{sec:overview_hotenets}).

We devise mechanism that utilizes the feature importance, locality/importance of the nodes (number of incoming and outgoing edges, availability of redundant paths for this node, and location of the node) on which the bugs occurred and using fusion technique to calculates severity of each bug and ranks them according to their severity. 

\item
We evaluate the efficacy of our tool (\maverick) with four different medium -- large scale campus and enterprise networks with each medium network infrastructures having upto 220 network nodes and large-scale network having 450 nodes, which includes routers, firewalls, switches, proxies, and gateway nodes (see \S \ref{sec:evaluation_hotnets}).

\end{packeditemize}


\section{Background \& Motivation}\label{sec:motivation_hotnets}

\blue{With increase in the scale, and lack of unified programming specification interfaces for enterprise and cloud networks makes network verification a challenging task. Today, majority of the network administrators still rely on plain-text configuration templates, command-line utilities, and wide variety of vendor-supplied programming specifications or user-interfaces for programming their networks~\cite{juniper_configuration_templates, cisco_configuration_templates, google_cloud_status_dashboard, 451research_network_automation_challenges}. This results in administrators unintentionally introducing bugs in the network configurations resulting in network outages or leaves network vulnerable to attacks~\cite{top_reason_network_outage_human_error_1, biggest_risk_uptime_huma_error2}.} 

\vspace{3pt}
\noindent \blue{\textbf{Configurations clusters}. We understand that for programming the network and creating policies requires same set of rules to be specified on wide range of devices that are present with in a network. Consider for example an ACL that is specified to allow {\tt TCP} traffic that is destined to WAN network {\tt 100.100.100.0/24} on port {\tt 1400} requires bunch of same ACLs to be specified on multiple routers or firewalls along multiple paths in which the traffic traverses. 
Similarly,  route-map entries, NAT rules, and route-filters specific to this ACL are also required to be specified on these routers  along the paths in which the traffic traverses. In general,  administrators either use sample templates or use CLI to configure multiple routers, which might result in human introduced errors.}

\begin{figure}[t!]
\begin{subfigure}[t]{0.23\textwidth}
\includegraphics{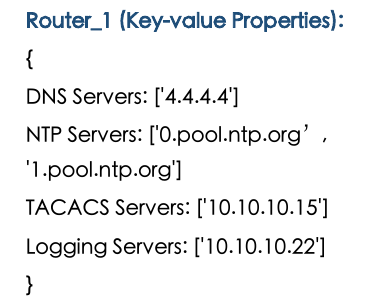}
\caption{Server properties (Key-value/s).}\label{fig:key_value}
\end{subfigure}\hfill
\begin{subfigure}[t]{0.23\textwidth}
\includegraphics{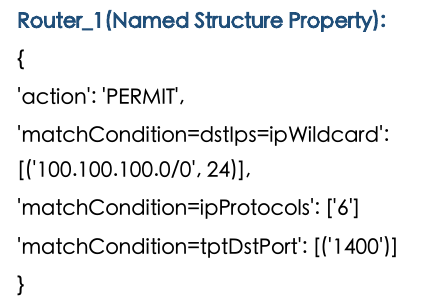}
\caption{IP ACL (Named structure).}\label{fig:named_structure}
\end{subfigure}
\caption{Illustrating {\em key-value} and {\em named-structure} properties in network configurations. {\em Named-structures} have complex nested key-value pairs. For brevity, the named-structure property (IP ACL) is made concise in the above example.}\label{figure:key_values_named_structures}
\end{figure}

\blue{We broadly classify overall network configurations in to two property classes (Figure \ref{figure:key_values_named_structures}): ($i$) {\em Key-value properties}, and ($ii$) {\em Named-structure properties}. As illustrated in Figure \ref{fig:key_value}, {\em key-value} property is a simple key:value/s pair that represents a discrete and independent network configuration (e.g., NTP Server configured for Router\_1 in Figure \ref{fig:key_value}). While the {\em named-structure} properties are structures with multiple key:value pairs nested as a complex discrete entity required to configure the network.} 


\subsection{Problems with Existing Approaches}\label{sec:problems_existing_approaches_hotnets}
\blue{For effectively detecting the bugs present in the networking configurations, the approaches aim to supplement the manual effort of network administrators by flagging probable network configuration and  data plane bugs~\cite{survey_network_configuration_errors,survey_network_verification_2}.}  These approaches broadly fall into two categories: ($a$) Statistical approach, and ($b$) logical or rule-based approach.

\vspace{3pt}
\noindent \textbf{Statistical techniques.} Statistical approaches aim to identify outliers in the configurations, and flag them as probable bugs~\cite{misconfigurations_intentionet_statistical, statistical_anomaly_detection, engler_2003_z-ranking, engler_2004_correlation}. For example, we illustrate the use of z-score, modified z-score and Gaussian mixture model (GMM) in Figure \ref{figure:motivation_zscore_mzscore_gmm}.
\blue{Though, we note that the outlying configurations have a much higher probability of being bugs, but, that in itself is not sufficient to detect the real bugs and highlight its severity.} The key disadvantages of such statistical approaches are as follows:

\begin{packeditemize}

    \item {\em High mis-classification rate}: Since many of the configurations lie at the boundary of the threshold used to classify as bugs, a large number of either false positives (i.e. incorrectly flagged as bugs) or false negatives (i.e. incorrectly flagged as valid) are identified. Correctly identifying the actual bugs from these lists again requires a lot of manual effort on the part of the administrator (See Figure \ref{figure:motivation_zscore_mzscore_gmm}).
    
    \item {\em Flagging intentional configuration changes}: Administrators might intentionally change configurations in specific ways to handle an uncommon use case. However, statistical techniques identify even such changes as configuration bugs (See Figure \ref{figure:motivation_zscore_mzscore_gmm}).
    
    \item {\em Critical bugs vs false positives}: In general, not all network configuration bugs are equally critical. Some bugs require immediate attention from administrators, whereas other bugs can be fixed slowly. However, we lack mechanism to identify the bugs that are critical in nature.
    
\end{packeditemize}

\vspace{3pt}
\noindent \textbf{Logical or rule-based techniques.} This approach is to let users specify grammar rules, any violation of such grammar rules is flagged as a configuration bug \cite{fogel2015_generalapproach_batfish, beckett2017general_minesweeper, yin_misconfig_errors_1, yin_misconfig_errors_2}. However, this approach too suffers from a number of drawbacks:


\begin{figure*}[tbp!]
\begin{center}
\centering
\includegraphics[width=0.8\linewidth]{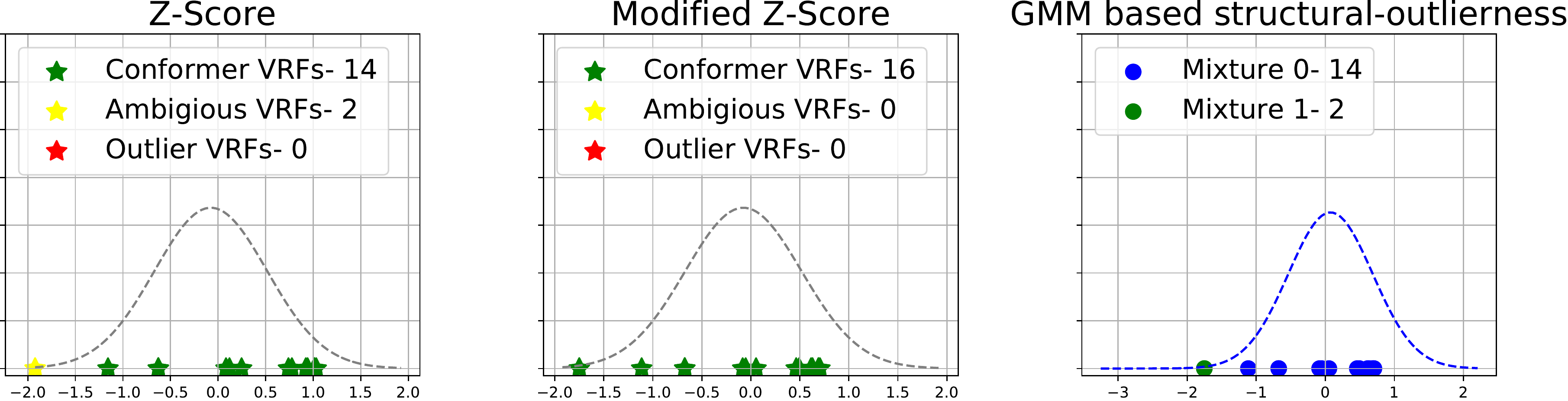}
\vspace{-5pt}
\includegraphics[width=0.8\linewidth]{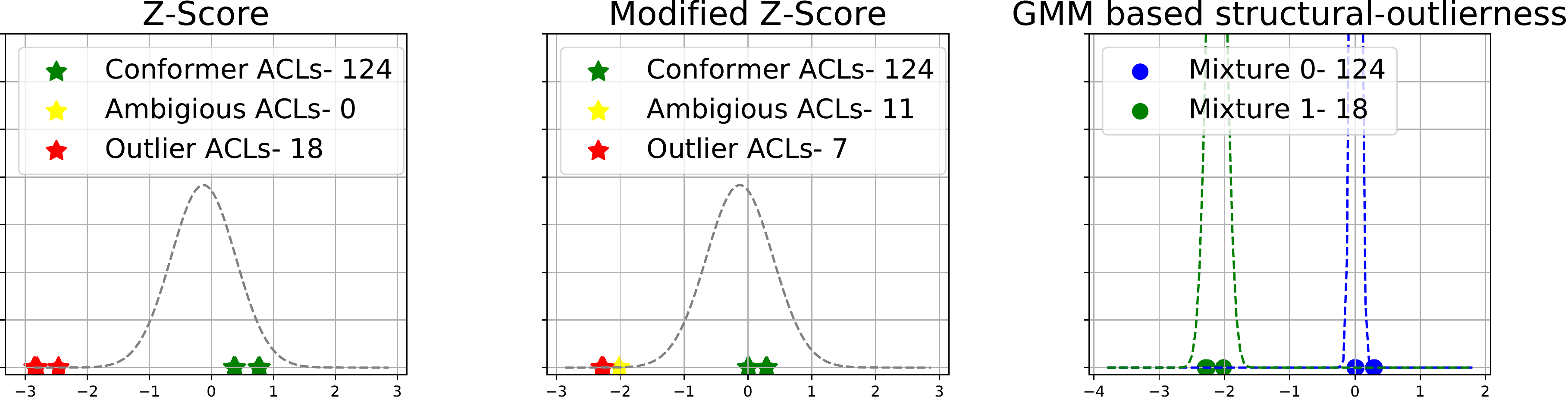}
\caption{Bug detection using statistical approaches such as z-score, modified z-score and GMM for VRF's and ACLs of network ({\tt DS-1}).}\label{figure:motivation_zscore_mzscore_gmm}
\end{center}
\end{figure*}

\begin{packeditemize}

    \item {\em Requirement of low-level specification rules}: It requires administrators, who are not always technically proficient, to specify complex low-level grammar rules. Thus, this is usually a cumbersome and technically involved process, that is also prone to mistakes.
    
    \item {\em Lack of coverage}: Even for technically proficient administrators, it is challenging to anticipate all the types of valid or invalid configurations and specify them in advance. Thus, many configuration bugs may pass through without getting identified.
    
    \item {\em Dependency on vendor-specific language}: Since each vendor uses different syntax to specify configurations, such vendor-specific syntax introduces an additional amount of complexity in specifying rules. Some tools \cite{fogel2015_generalapproach_batfish} try to handle this complexity by converting all the configuration files into vendor-independent language, but generalization can lead to increase in the number of unidentified bugs.
\end{packeditemize}

Therefore, it is becoming increasingly difficult to proactively detect the network configuration bugs i.e., before deploying them on to the production networks.

\section{MAVERICK Overview}\label{sec:overview_hotenets}

\begin{figure}[t!]
\centering
\includegraphics{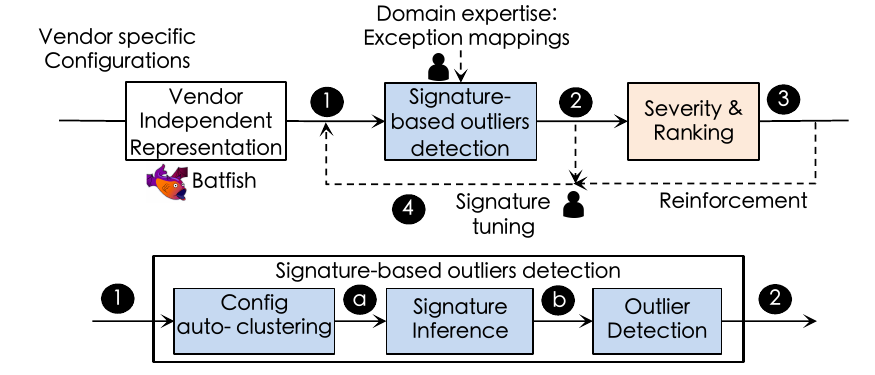}
\caption{\maverick System Architecture.}\label{figure:highlevel_system_arch_outliers}
\vspace{-3pt}

\end{figure}

\blue{We present the overview of \maverick control plane network verification engine that is a tangible step toward addressing the limitations discussed above (\S \ref{sec:problems_existing_approaches_hotnets}).  Figure \ref{figure:highlevel_system_arch_outliers} provides an overview of the \maverick system architecture, with the following two key capabilities: ($i$) {\em signature-based outlier detection} engine, and ($ii$) {\em severity \& ranking} engine. These capabilities allows administrators to proactively detect the control plane bugs and fine-tune them to reduce the false positives, while reducing their time vested in triaging critical bugs rather than spending time on false-positives.}

\vspace{3pt}
\noindent {\bf Signature-based outlier detection}. This module digests the total network configurations (i.e., provided in {\em vendor-independent} specification\footnote{We use batfish~\cite{fogel2015_generalapproach_batfish} for generating the vendor-independent specification. Batfish presents capability to digest the network configurations presented in various vendor-specific formats and translate them into vendor-independent specifications.}) for detecting the bugs (\ballnumber{1}). We use {\em structural outlierness} as fundamental ingredient used to effectively detect bugs in the network configurations. We define, {\em structural outlierness} as the deviation of a network property (i.e., key-value property or named-structure) from its group or cluster of configurations (i.e., most popular entries of the cluster) that are programmed onto multiple nodes to achieve the same functionality (as discussed in \S \ref{sec:motivation_hotnets}) (\ballnumber{a}). For the derived most popular entries within a group or cluster, we apply additional heuristics and domain expertise (i.e., captured as {\em exception mappings}) for automatically inferring the signatures (\ballnumber{b}). We supply such automatically inferred signatures to administrator for inspection and fine-tuning these signatures for detecting bugs in the network configurations (\ballnumber{2}). Though administrator's intervention is optional in our case, we use {\em human-in-loop} for reducing false positives (\ballnumber{4}).) These signatures we inferred could be used to detect bugs in the network configurations before applying them to the network ({\em Signature-based outlier detection}).

The capabilities that are discussed above are performed by following three key modules of {\em signature-based outlier detection} engine (details discussed in \S \ref{sec:design_hotnets}): ($a$) {\em Config auto-clustering} module, ($b$) {\em signature-inference} engine, and ($c$) {\em Outlier detection} engine. 


\noindent {\bf Severity \& ranking}. We apply the severity and ranking mechanism we developed to re-prioritize the bugs for effectively identifying their severity, which helps reduce administrator's effort and time spent in handling false positives (\ballnumber{3}). We use following three key metrics for calculating the severity for ranking the outliers (details discussed in \S \ref{ref:severity_ranking}): ($a$) {\em Similarity and outlierness scores}, ($b$) Well connected-ness of nodes, ($c$) {\em Feature-dependency}.

\noindent \textbf{Design goal.} Our goal is to mitigate the problems in existing enterprise and campus networks by reducing the amount of effort involved in detecting bugs, automatically inferring signatures that acts as reference to verify the network configurations for its sanity and bugs, while increasing the network coverage (for detecting generic bugs). \blue{Unlike, existing techniques which requires network configurations to be manually grouped for building the templates~\cite{misconfigurations_intentionet_statistical}, \maverick automatically clusters the configurations into separate groups for building the signatures.} However, a key drawback of such signature inference is that it falsely flags configurations that network administrators have designed for customized use cases. To mitigate this problem, we allow administrators to re-tune inferred signatures. Since there can be multiple valid signatures, this also automatically allows more customized configurations.

We recognize that even with multiple signatures, it is possible to false classify multiple configurations as bugs. However, not all configuration bugs are equally important in a network. \blue{Based on the anticipated impact of bugs, we evaluate the severity and assign a priority score to each of the identified bugs and rank them accordingly.} This allows the administrators to focus on the most important bugs, while letting the less important ones remain for longer time.

\section{System high level design}\label{sec:design_hotnets}

In this section, we discuss the network verification engine that we developed to address the limitations discussed above in \S \ref{sec:problems_existing_approaches_hotnets}. As shown in Figure \ref{figure:highlevel_system_arch_outliers}, \maverick supports following key functional components to address these challenges: ($i$)  Vendor-independent (VI) representation of network configurations, ($ii$) Signature-based outlier detection, and ($iii$) Severity and ranking mechanism.

\subsection{VI Representation \& Encoding}
We use Batfish~\cite{fogel2015_generalapproach_batfish} to translate the network configurations from vendor-specific languages (e.g., Cisco's IOS, Juniper's JunOS) to vendor-independent (VI) representation. This avoids the need for designing parsers for each of the vendor-specific language in \mavericknospace. 
We extract each of the named structures (such as ACL's, route-maps, route-policies), server and interface properties (such as DNS server, NTP server configurations on all the network devices) and encode such categorical data into binary encoded format i.e., using Multi-label binarizer ~\cite{categorical_data_encoding}, which allows us to apply statistical and machine learning (ML) techniques on the network configuration data.

\subsection{Signature-based Outlier Detection}
The key challenge in bug detection is the ability of administrator to craft the specification or signature that allows the tool to detect the bugs and errors. Therefore, automatically generating (i.e., inferring) the signatures is the key step towards effective detection of bugs in the network configurations. \mavericknospace's signature-based outlier detection engine supports following three key capabilities for automatically detecting the bugs present in the network configurations represented in VI format.

\subsubsection{Configuration auto-clustering} 

As a first step, we run clustering on each of the named structures (such as ACLs, router-filters, route-maps) independently, to segregate them on the basis of their categories and properties. For example, a network with thousands of ACLs are clustered into group of tens or groups of hundred on the basis of their similarity for automatically inferring signatures from each of the ACL groups, which is required to compute its signature. As manually grouping thousands of ACLs into groups on the basis of their name or other properties is a challenging and tedious process,  
 we use K-means a ML-based technique to cluster the named structures. The clustered named structures are then used for signature inference. To obtain the right value of K, we use Elbow technique \cite{elbow1} and regress on different values of K to decide the optimum. \blue{We heuristically choose a lower limit of K equal to the number of unique set of names used to configure different named structures.  Therefore, clustering reduces the number of signatures inferred, thereby reducing the amount of manual effort involved with administrator in verifying the signatures to re-tune them for increasing the precision of signature-based outlier detection.} 


\begin{table*}[t]
\resizebox{\textwidth}{!}{%
\begin{tabular}{llllllll}
\hline
\multicolumn{1}{|c|}{\textbf{Outlier}} &
  \multicolumn{1}{c|}{\textbf{Signature Definition}} &
  \multicolumn{1}{c|}{\textbf{\begin{tabular}[c]{@{}c@{}}Conformer\\    Nodes\end{tabular}}} &
  \multicolumn{1}{c|}{\textbf{Outlier Definition}} &
  \multicolumn{1}{c|}{\textbf{\begin{tabular}[c]{@{}c@{}}Outlier\\  Nodes\end{tabular}}} &
  \multicolumn{1}{c|}{\textbf{\begin{tabular}[c]{@{}c@{}}Outlier\\ Properties\end{tabular}}} &
  \multicolumn{1}{c|}{\textbf{\begin{tabular}[c]{@{}c@{}}Outlierness\\     Value\end{tabular}}} &
  \multicolumn{1}{c|}{\textbf{\begin{tabular}[c]{@{}c@{}}Severity\\   Score\end{tabular}}} \\ \hline
\multicolumn{1}{|l|}{\begin{tabular}[c]{@{}l@{}}outlier:Route\_\\ Filter\_List\_0\end{tabular}} &
  \multicolumn{1}{l|}{\begin{tabular}[c]{@{}l@{}}\{'action': {[}{[}'PERMIT', 16{]}{]}, \\ 'ipWildcard':{[}{[}'100.100.100.0/23', '*', 9{]},   \\   ..........................             {[}'25-25', '*', 10{]}{]}\}\end{tabular}} &
  \multicolumn{1}{l|}{\begin{tabular}[c]{@{}l@{}}{[}'rt1-na-dc1', \\ 'rt2-na-dc1'.\\ .., rt91-na-dc1{]}\end{tabular}} &
  \multicolumn{1}{l|}{\begin{tabular}[c]{@{}l@{}}\{'action': 'PERMIT', 'ipWildcard': \\ '100.100.0.0/16', 'lengthRange': '16-20'\}\end{tabular}} &
  \multicolumn{1}{l|}{\begin{tabular}[c]{@{}l@{}}{[}'rt19-na-dc1', \\ 'rt28-na-dc1'{]}\end{tabular}} &
  \multicolumn{1}{l|}{\begin{tabular}[c]{@{}l@{}}{[}{[}'lengthRange', \\ '16-20'{]}{]}\end{tabular}} &
  \multicolumn{1}{l|}{0.978} &
  \multicolumn{1}{l|}{1.177} \\ \hline
 &  &  &  &  &  &  &  \\
 &  &  &  &  &  &  & 
\end{tabular}}
\caption{Final ranked bug outcome \maverick tool in accordance with its severity.}
\label{tab:output_maverick_tool}
\end{table*}

\subsubsection{Signature Inference \& Generalization}\label{sec:signature_inference}


The signature inference engine automatically infers and builds the signatures from the clustered named structures. 
The signature inference engine composes all the named part of the cluster to frame a single signature. We use following grammar in our signature for effectively capturing and generalizing the signatures, which includes following operators: `*', `!' `=' `[]', `\{\}', `OR', `AND', <IP-Subnet> (i.e., IP specific to that subnet will be considered as legitimate in the signature).


\begin{figure}[t!]
\centering
\includegraphics[width=0.45\textwidth]{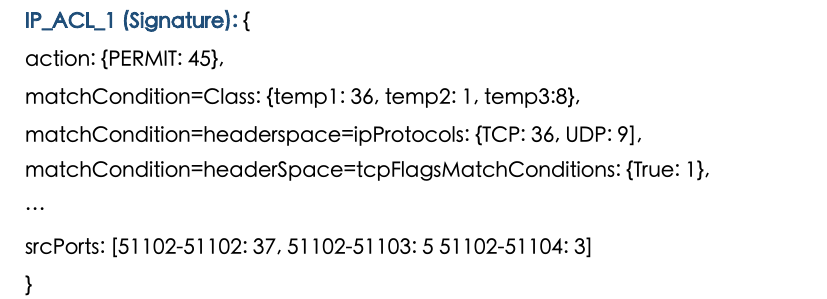}
\caption{Signature Inferred by \maverick for IP ACL withits popularity weights are shown above. Only part of the signature is shown for brevity.}\label{figure:inferred_signatures}
\end{figure}

As shown in Figure \ref{figure:inferred_signatures}, the signature of a named structure includes set of key-value pairs (i.e., complex nested). The Key is property name and the values are array of tuples. The tuples captures one of the values of property and its weight, where as weight represents the frequency of occurrences or the density of the value for that property with in that cluster. 


\begin{algorithm}[t!]
\footnotesize
\caption{Signature Inference Algorithm.}\label{table:signature_inference_algo}

\begin{algorithmic}[1]
\State $F \gets {generateVI()}$
\State {$P \gets $getNamedStructProps($F$)}
\State {$P \gets $encode($P$)}
\State {$K \gets$ elbow($P$)}
\State {$C \gets$ clusters using K-Means of $P$}
\State {Let $F(c,p)$ be the value-frequency pair $\forall c\in C, p \in P$.}
\State {Compute threshold $T(c, p)$ from $F(c, p)$, $\forall c \in C, p \in P$. }
\State {Let $\epsilon$ be the margin of uncertainty}
\For {$c \in C$}
    \For {$p \in P$}
    \For {$(k,v) \in F(c, p)$}
        \If {$v > T(c,p) + \epsilon$}
        \State {Mark $p$ as bug}
        \ElsIf{$v > T(c,p) \And v<T(c,p) + \epsilon$}
        \State {Mark $p$ as probable bug}
        \Else
        \State {Mark $p$ as normal property}
        \EndIf
        \EndFor
    \EndFor
\EndFor

\end{algorithmic}
\end{algorithm}

Also, the ability of these techniques to effectively accommodate the domain expertise and inputs from administrators allows them to effectively detect bugs present in the network. The signature-mappings enforces constraints on the property's key:value pairs that are part of the signature. The signature-mapping which is provided as the domain knowledge from the administrator restricts the signature inference engine to treat specific key:value pairs differently. 
For example, the inference engine discards \blue{any specific key `XYZ' from being part of the signature and key `YYZ' to be considered as mandatory. This allows us to {\em white-list}, create {\em exception}, or {\em black-list} specific keys to the signature inference engine about the way it should consider the respective key:value pairs.}


\subsubsection{Re-tuning \& Outlier detection}
\blue{Signatures auto-generated using ML-based techniques could be further fine-tuned by administrator by supplying the domain knowledge as {\em signature-mappings} 
or manual inspection. On the contrary, for simple server properties (e.g., DNS servers, TACACS server properties) the names used on different nodes are required to be same, which simplifies our task of grouping configurations for clustering to detect outliers. Hence, they could be simply grouped together for calculating the outliers.}

To verify if a named structure is an outlier, we compare the properties of this named structure with the respective properties of the cluster signature. If all the properties in the named structure that is compared with the signature matches, then the named structure is considered as valid and bug otherwise. 
We also calculate their similarity scores $S_i$ and outlier scores $O_i$, to determine the amount by which a named structure matches with the signature.

\vspace{-0.05in}
\begin{equation}
\label{equation:similarity_score}
S_i =  \frac{\sum_{i=1}^n W_{i}}{\sum_{j=1}^s W_{j}}, O_i=1-S_i, \, \forall i = 1, \ldots, n; \ \forall j = 1, \ldots, s,
\end{equation}
\blue{where s is total number of properties in the signature, $W_i$ represents the weight associated with each of the property in the signature and s is total number of signatures.}
\vspace{-0.05in}



\subsection{Severity and Ranking}\label{ref:severity_ranking}
\begin{figure}[t!]
\centering
\includegraphics[width=1.0\linewidth]{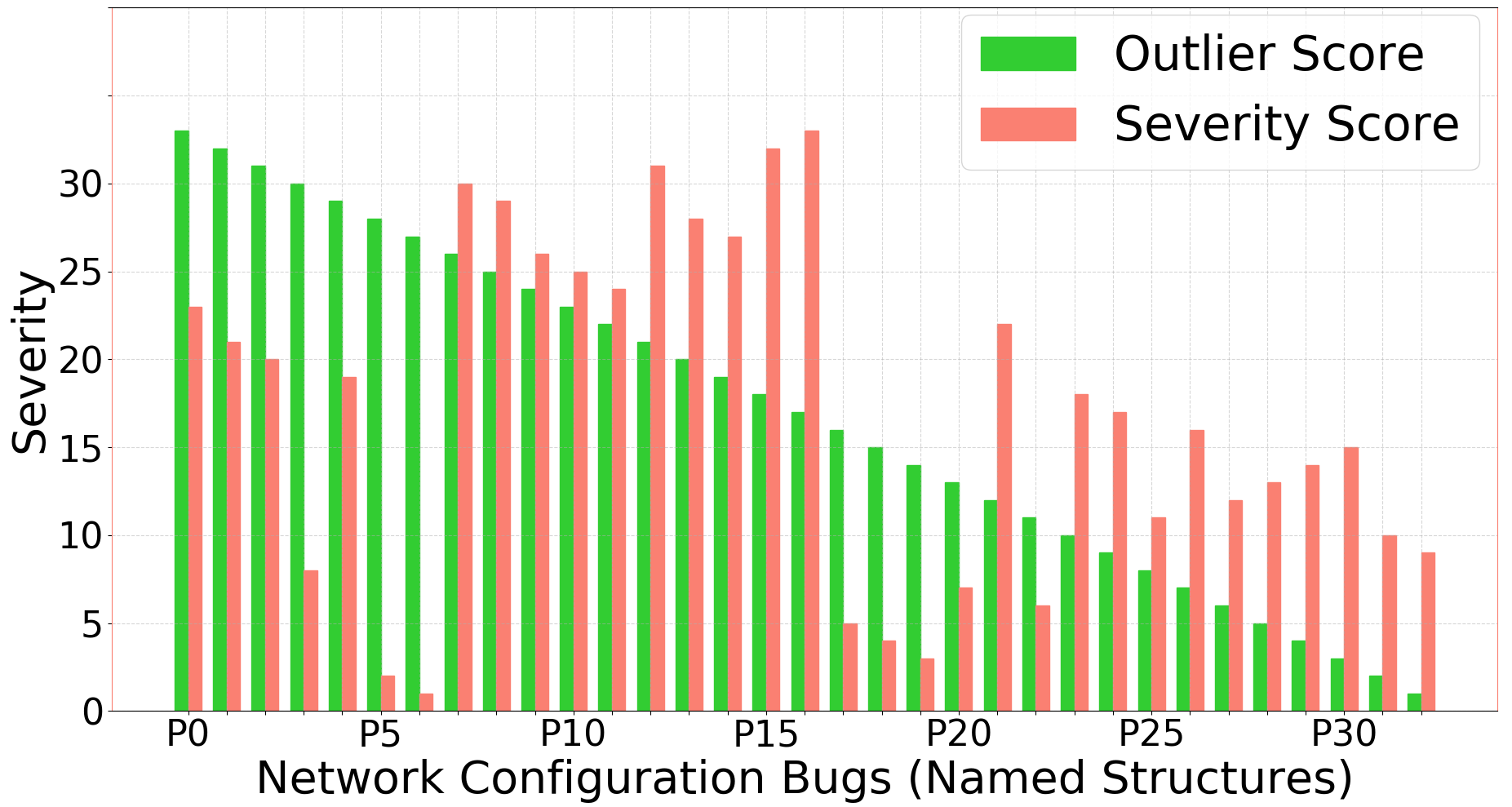}
\caption{Bugs discovered by \mavericknospace's with and without severity applied.}
\label{figure:bug_severity}
\end{figure}

This list of outliers that is generated as outcome of the signature-based outliers engine contains the outlier definition, the named structure it belongs to, and it's outlier score. The outlier score is an indication of how strongly our engine believes a particular outlier to be a bug and its value is between 0 and 1. But an entry with a very high score could mean that it is a single separate configuration and does not belong to any signature. Our severity and ranking mechanism takes this into consideration for effectively calculating the severity. To rank these outliers, we devise different metrics and assign each outlier a metric score. Then, using a particular combination of these metric scores, we calculate the final score of each outlier and rank them based on this score. \maverick uses three different metrics to calculate the severity and ranking of the outliers: 
\begin{packeditemize}

\item[1.] {\em Similarity and outlierness scores} that we derived from the outcome of signature-based outlier engine is used as one factor in deciding the severity of the final bug outcome.

\item[2.]  {\em Well-connectedness} of nodes: We use the page-rank algorithm to establish the importance of each node with the general idea being that a possible bug in a more important or well-connected node would be more severe than a bug that has fewer connections.  

\item[3.] {\em Feature-Dependency Score}: This metric tells us the importance of the features that the named structure is a part of. The general idea is that the importance of named-structure is network-specific and therefore, dynamically evaluating these scores helps provide a much finer and network-specific bug severity analysis. Consider for example, when a ACL rule marked as outlier will results in impacting the NAT rules, route-filters and VRFs associated with it. Hence, outliers in features that has higher dependency with other features will result in high severe bugs. The final outcome of the severity and ranking module results in generating bugs that result in lesser in FPs and FNs (Table \ref{tab:precision_recall}) and effectively ranked according to its severity (Figure \ref{figure:bug_severity}).

\end{packeditemize}

Finally, the {\em human-in-the-loop} correlation Score helps re-tune the signature and reduces false-positives. Once the Network Administrator flags a certain Outlier as a Bug or a FP, all the corresponding Outliers in the population (i.e., cluster in our case) show an increase or a decrease in their severity score respectively. This metric allows the administrator to manually inspect numerous bugs of a specific type from a very large network with relative ease.




\section{Prototype Evaluation}\label{sec:evaluation_hotnets}





\begin{figure*}[h!]
\centering
\includegraphics[width=0.9\linewidth]{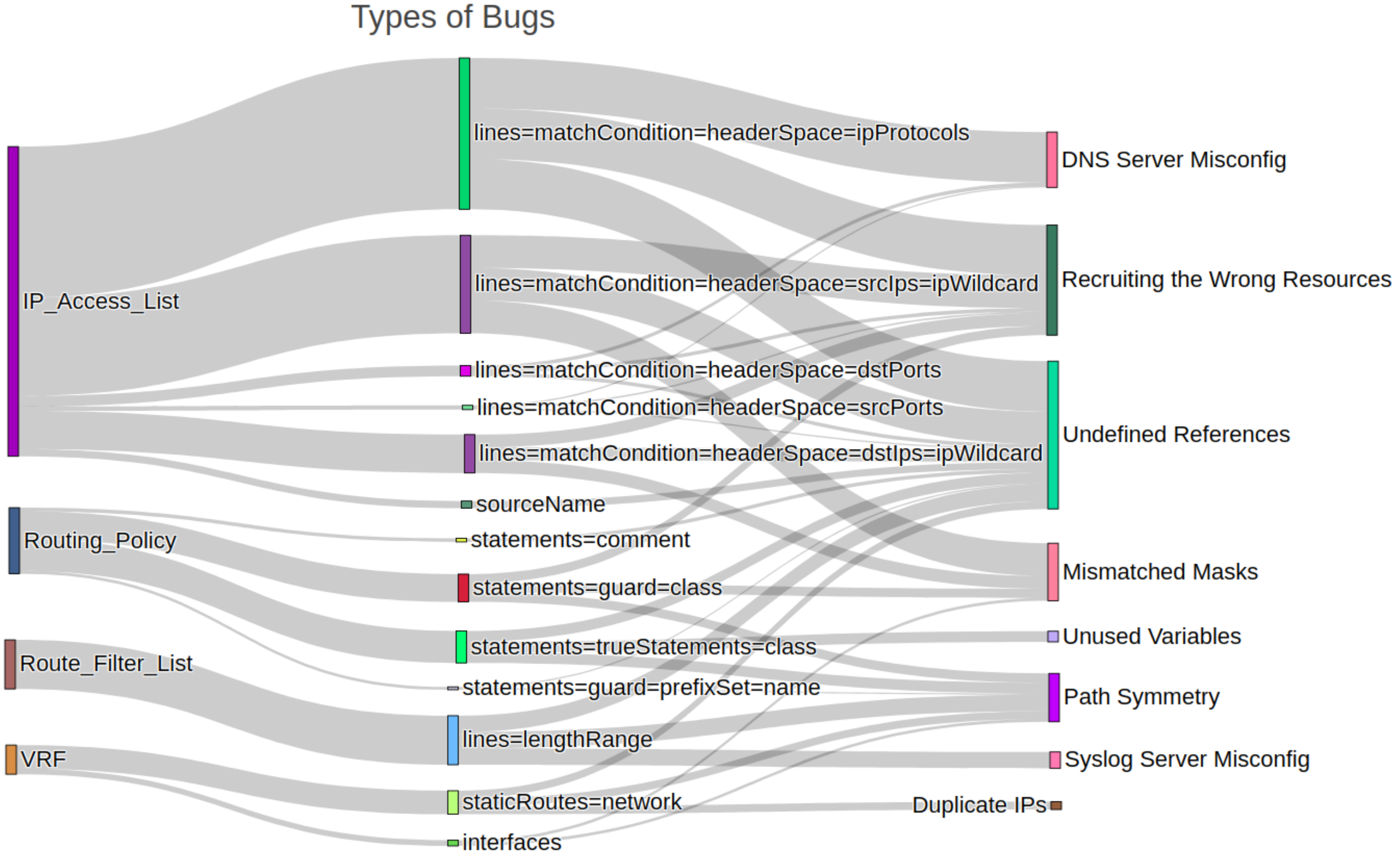}
\caption{Correlation of bugs discovered by \maverick with real-world network problems. The left layer corresponds to the type of property in which outlier is detected. The middle layer corresponds to the type of signature that is violated. The right layer is the type of real-world network problem. A thicker flow denotes a stronger correlation.}
\label{figure:bug_type_plot}

\end{figure*}

\noindent \textbf{Dataset: }
We evaluate the performance of \maverick over a total of four networks using their configuration files.
Of the four networks, three are of medium size network of 157, 132 and 221 nodes and large network of 454 nodes. Medium networks has around 5000 -- 10000 properties, while large scale network has around 60000 properties. The properties consist of ACLs, Route Filters, VRFs and Routing Policies with ACLs predominant in the large network, whereas RouteFilters are predominant in the medium sized networks.




\begin{table}[!th]
  \footnotesize
  \centering
\begin{tabular}{|p{0.95in}|p{0.2in}|p{0.2in}|p{0.2in}|p{0.4in}|p{0.4in}|}
\hline
\Centering\bfseries \textbf{Approach} & \Centering\bfseries \textbf{TP} & \Centering\bfseries \textbf{FP} & \Centering\bfseries \textbf{FN} & \Centering\bfseries \textbf{Precision} & \Centering\bfseries \textbf{Recall} \\ \hline

Z-score & 392  &  1031  & 240  & 0.275 & 0.620  \\ \hline
Modified Z-score & 417 &  692  & 132  & 0.386  & 0.760 \\ \hline
GMM &  298  &  608  & 220  & 0.329  & 0.575 \\ \hline
Maverick (Outliers)   &  472 &  154  & 32  & 0.754  & 0.937  \\ \hline
Maverick (Retuning)    & 498 & 32  &  8  & 0.940   & 0.984                        \\ \hline            
\end{tabular}
\vspace{5pt}
\caption{Efficacy of the \maverick for medium scale enterprise network dataset ({\tt \textbf{DS-3}}).  Reported outliers with each of the technique: Z-score: 1798, Modified z-score: 1296, GMM: 1157, Maverick (Outliers): 754, Maverick (Retuning): 574. Careful retuning of $\approx$97 clusters detected by \maverick required less than 2 hours for manual inspection.}
\end{table}

\subsection{Performance of \maverick}
We first compare the performance of \maverick in terms of precision and recall with baseline techniques (Table 2).
The baseline techniques include Z-score, modified Z-score, GMM, and \maverick using only outliers.
We make two major observations.
First, we note that \mavericknospace's outlier detection performs better than Z-score, modified Z-score and GMM in terms of both precision and recall.
However, the precision is still only around 0.754, which has further scope of improvement.
This is primarily due to presence of false positives, as a large number of outliers are detected using the inferred signatures.
Second, manual retuning of signatures can then further increase the precision to 0.94, thus increasing by 19\% compared to just outlier-based detection.
This shows that manual retuning of signatures can significantly improve the precision.

\subsection{Importance of Severity Score}
We now look at how severity score can change the sequence of bugs shown to administrators (Figure \ref{figure:bug_type_plot}).
We plot the bugs reported in the sequence of their outlier score, along with their severity scores for one of the medium-sized network.
We note that the sequence of bugs shown to the administrators changes considerably, with P16 rising up to the most severe rank, followed by P15 and P13.
On the other hand, P6 reduces to the least severe rank, followed by P5.
This shows that using severity score alters the sequence of bugs shown to the users, and can lead to less important bugs being given less priority even if they have high outlier scores.

\subsection{Correlation of bugs with real network issues}
To further observe the type of bugs \maverick discovers, we utilize a sankey diagram to show how outliers in different properties correspond to different types of bugs (Figure \ref{figure:bug_type_plot} in one of the medium-sized network.
A higher thickness of flow denotes a higher number of bugs corresponding to a specific signature in the middle layer of vertices and then to types of network problems in the last layer.
We observe that most of the bugs arise due to problems in IP access lists, followed by routing policy, route filter list and VNF's.
We also observe that the most common type of network problem is undefined references, but each type of outlier roughly has equal probability of leading to an undefined reference.
In the future, we plan to further study if the type of network problem can also be inferred using this statistics.





\section{Conclusion}\label{sec:conclusion_hotnets}

\noindent This paper presented a novel signature inference framework for detecting the control plane bugs based on structural deviations (i.e., outliers or bugs), while their severity is estimated and bugs are ranked accordingly. The key strength of this work lies in its ability to automatically infer the signatures from raw network configurations without much administrator's intervention and generalize these inferred signatures for transportability. We combine disparate metrics to rank the severity of the detected outliers. We evaluated our approach using four different datasets of campus networks and achieved high bug detection of up to 97\% with supply of domain expertise in the form of signature-mappings. While our approach was simple, with inferred signatures we were able to discover numerous bugs, including those that would impossible to discover with existing network validation tools.

\bibliographystyle{abbrv} 
\bibliography{HotNets/reference}

\begin{thebibliography}{10}

\bibitem{gartner_downtime_cost}
{The Cost of Downtime.}
\newblock
  \url{https://blogs.gartner.com/andrew-lerner/2014/07/16/the-cost-of-downtime/},
  July 2014.

\bibitem{juniper_configuration_templates}
{Use templates to define a common device configuration}.
\newblock {\em Product Documentation}, 2017.

\bibitem{cisco_configuration_templates}
{Managing Multiple Networks with Configuration Templates}.
\newblock {\em Product Documentation}, 2018.

\bibitem{ugly_truth_downtime_cost}
{The Ugly Truth about Downtime Costs and How to Calculate Your Own.}
\newblock \url{https://www.itondemand.com/2018/05/29/costs-of-downtime/}, May
  2018.

\bibitem{google_cloud_status_dashboard}
{Google Cloud Networking Incident 19009}.
\newblock {\em Google Cloud Networking Incidents}, 2019.

\bibitem{451research_network_automation_challenges}
{451 Research: Enterprise Network Automation Gets Competitive}.
\newblock {\em Technical Report}, 2020.

\bibitem{beckett2017general_minesweeper}
R.~Beckett, A.~Gupta, R.~Mahajan, and D.~Walker.
\newblock A general approach to network configuration verification.
\newblock In {\em Proceedings of the Conference of the ACM Special Interest
  Group on Data Communication}, pages 155--168, 2017.

\bibitem{top_reason_network_outage_human_error_1}
A.~Bednarz.
\newblock {Top reasons for network downtime: Network outages linked to human
  error, incompatible changes, greater complexity}.
\newblock {\em Technical Report}, 2018.

\bibitem{elbow1}
P.~Bholowalia and A.~Kumar.
\newblock Ebk-means: A clustering technique based on elbow method and k-means
  in wsn.
\newblock {\em International Journal of Computer Applications}, 105:17--24,
  2014.

\bibitem{statistical_anomaly_detection}
C.~Callegari, S.~Vaton, and M.~Pagano.
\newblock A new statistical approach to network anomaly detection.
\newblock In {\em 2008 International Symposium on Performance Evaluation of
  Computer and Telecommunication Systems)}, pages 441 -- 447, 2008.

\bibitem{fogel2015_generalapproach_batfish}
A.~Fogel, S.~Fung, L.~Pedrosa, M.~Walraed-Sullivan, R.~Govindan, R.~Mahajan,
  and T.~Millstein.
\newblock A general approach to network configuration analysis.
\newblock In {\em 12th $\{$USENIX$\}$ Symposium on Networked Systems Design and
  Implementation ($\{$NSDI$\}$ 15)}, pages 469--483, 2015.

\bibitem{arc}
A.~Gember-Jacobson, R.~Viswanathan, A.~Akella, and R.~Mahajan.
\newblock Fast control plane analysis using an abstract representation.
\newblock In {\em Proceedings of the 2016 ACM SIGCOMM Conference}, pages
  300--313. ACM, 2016.

\bibitem{misconfigurations_intentionet_statistical}
S.~K.~R. Kakarla, A.~Tang, R.~Beckett, K.~Jayaraman, T.~Millstein, Y.~Tamir,
  and G.~Varghese.
\newblock {Finding Network Misconfigurations by Automatic Template Inference}.
\newblock In {\em 17th {USENIX} Symposium on Networked Systems Design and
  Implementation ({NSDI} 20)}, pages 999--1013, Santa Clara, CA, Feb. 2020.
  {USENIX} Association.

\bibitem{khurshid2013veriflow}
A.~Khurshid, X.~Zou, W.~Zhou, M.~Caesar, and P.~B. Godfrey.
\newblock Veriflow: Verifying network-wide invariants in real time.
\newblock In {\em Presented as part of the 10th $\{$USENIX$\}$ Symposium on
  Networked Systems Design and Implementation ($\{$NSDI$\}$ 13)}, pages 15--27,
  2013.

\bibitem{engler_2004_correlation}
T.~Kremenek, K.~Ashcraft, J.~Yang, and D.~Engler.
\newblock Correlation exploitation in error ranking.
\newblock In {\em ACM SIGSOFT Software Engineering Notes}, volume~29, pages
  83--93. ACM, 2004.

\bibitem{engler_2003_z-ranking}
T.~Kremenek and D.~Engler.
\newblock Z-ranking: Using statistical analysis to counter the impact of static
  analysis approximations.
\newblock In {\em International Static Analysis Symposium}, pages 295--315.
  Springer, 2003.

\bibitem{survey_network_verification_2}
Y.~{Li}, X.~{Yin}, Z.~{Wang}, J.~{Yao}, X.~{Shi}, J.~{Wu}, H.~{Zhang}, and
  Q.~{Wang}.
\newblock A survey on network verification and testing with formal methods:
  Approaches and challenges.
\newblock {\em IEEE Communications Surveys Tutorials}, 21(1):940--969, 2019.

\bibitem{panda2015new_directions}
A.~Panda, K.~Argyraki, M.~Sagiv, M.~Schapira, and S.~Shenker.
\newblock New directions for network verification.
\newblock In {\em 1st Summit on Advances in Programming Languages (SNAPL
  2015)}. Schloss Dagstuhl-Leibniz-Zentrum fuer Informatik, 2015.

\bibitem{biggest_risk_uptime_huma_error2}
A.~Patrizio.
\newblock {The biggest risk to uptime? Your staff: Human error is the chief
  cause of downtime, a new study finds. Imagine that.}
\newblock {\em Technical Report}, 2019.

\bibitem{categorical_data_encoding}
M.~A. Reddy.
\newblock Encoding categorical data in machine learning.
\newblock {\em Technical Article}, 2019.

\bibitem{yin_misconfig_errors_2}
T.~Xu, J.~Zhang, P.~Huang, J.~Zheng, T.~Sheng, D.~Yuan, Y.~Zhou, and
  S.~Pasupathy.
\newblock Do not blame users for misconfigurations.
\newblock In {\em Proceedings of the Twenty-Fourth ACM Symposium on Operating
  Systems Principles}, SOSP ’13, page 244–259, New York, NY, USA, 2013.
  Association for Computing Machinery.

\bibitem{survey_network_configuration_errors}
T.~Xu and Y.~Zhou.
\newblock Systems approaches to tackling configuration errors: A survey.
\newblock {\em ACM Comput. Surv.}, 47(4), July 2015.

\bibitem{yin_misconfig_errors_1}
Z.~Yin, X.~Ma, J.~Zheng, Y.~Zhou, L.~N. Bairavasundaram, and S.~Pasupathy.
\newblock An empirical study on configuration errors in commercial and open
  source systems.
\newblock In {\em Proceedings of the Twenty-Third ACM Symposium on Operating
  Systems Principles}, SOSP ’11, page 159–172, New York, NY, USA, 2011.
  Association for Computing Machinery.

\end{thebibliography}






\iffullpaper

\clearpage












\fi


\end{document}